\documentclass[12pt]{iopart}
\usepackage[T1]{fontenc}
\usepackage[latin9]{inputenc}
\usepackage{verbatim}
\usepackage{units}
\usepackage{textcomp}
\usepackage{setstack}
\usepackage{iopams}
\usepackage{graphicx}
\usepackage{esint}
\usepackage{color}

\makeatletter

\begin{document}

\title{Coupling single molecule magnets to quantum circuits}

\author{Mark Jenkins$^{1}$, Thomas H\"{u}mmer$^{2,3}$, Mar\'{\i}a Jos\'e Mart\'{\i}nez-P\'erez$^{1}$, Juanjo  Garc\'{\i}a-Ripoll$^{4}$, David Zueco$^{1,5}$, and Fernando Luis$^{1}$}

\address{$^{1}$ Instituto de Ciencia de Materiales de Arag\'on (CSIC-U. de Zaragoza) and Departamento de F\'{\i}sica de la Materia Condensada (U. de Zaragoza), $50009$ Zaragoza, Spain}
\address{$^{2}$ Ludwig-Maximilians-Universit\"at M\"unchen, D-80799 Munich, Germany}
\address{$^{3}$ Max-Planck-Institut f\"ur Quantenoptik, D-85748 Garching, Germany}
\address{$^{4}$ Instituto de F\'{\i}sica Fundamental, IFF-CSIC, Serrano $113$-bis, $28006$ Madrid, Spain}
\address{$^{5}$ Fundaci\'on ARAID, Paseo Mar\'{\i}a Agust\'{\i}n $36$, $50004$ Zaragoza, Spain}
\eads{mjenk@unizar.es,dzueco@gmail.com,fluis@unizar.es}

\begin{abstract}
In this work we study theoretically the coupling of single molecule magnets (SMMs) to a variety of quantum circuits, including microwave resonators with and without constrictions and flux qubits.  The main result of this study is that it is possible to achieve strong and ultrastrong coupling regimes between SMM crystals and the superconducting circuit, with strong hints that such a coupling could also be reached for individual molecules close to constrictions. Building on the resulting coupling strengths and the typical coherence times of these molecules ($\sim\mu$s), we conclude that SMMs can be used for coherent storage and manipulation of quantum information, either in the context of quantum computing or in quantum simulations. Throughout the work we also discuss in detail the family of molecules that are most suitable for such operations, based not only on the coupling strength, but also on the typical energy gaps and the simplicity with which they can be tuned and oriented. Finally, we also discuss practical advantages of SMMs, such as the possibility to fabricate the SMMs ensembles on the chip through the deposition of small droplets.
\end{abstract}

\submitto{\NJP}

\maketitle

\section{Introduction}
\label{Intro}

Solid-state spin ensembles are seen as promising media to store quantum information as well as to interconnect radio-frequency and optical photons \cite{Imamoglu2009,Wesenberg2009,Marcos2010}. Experiments performed in the last few years have shown the feasibility of coherently coupling NV or P1 centres in diamond to either superconducting resonators \cite{Schuster2010,Kubo2010,Amsuss2011} or flux qubits \cite{Zhu2011}. Evidences for strong magnetic coupling have also been found, even at room temperature, between spin-$1/2$ paramagnetic radicals and three-dimensional microwave cavities \cite{Chiorescu2010,Abe2011}. For such large ensembles, the magnetic coupling is enhanced with respect to the coupling of a single spin by a factor $\surd N$, where $N$ is the number of spins. Even more challenging is to coherently couple to individual spins. Provided this limit can be attained, on-chip superconducting circuits could be used to coherently manipulate and transfer information between spin qubits, thus providing a suitable architecture to implement an all-spin quantum processor \cite{Awschalom2013}.

In the present work, we consider a different family of magnetic materials: single molecule magnets (SMMs) \cite{Christou2000,Gatteschi2003,Bartolome2013}. These are organometallic molecules formed by a high-spin magnetic core surrounded by organic ligands that naturally organize into molecular crystals. In SMMs with strong uniaxial magnetic anisotropy, such as Mn$_{12}$ or Fe$_{8}$, the magnetization shows hysteresis (i.e. magnetic memory) near liquid Helium temperatures \cite{Sessoli1993}. In addition, SMMs show intriguing quantum phenomena such as resonant spin tunneling \cite{Friedman1996,Hernandez1996,Thomas1996,Sangregorio1997} and Berry phase interferences between different tunneling paths \cite{Wernsdorfer1999}.

SMMs are also attractive candidates to act as either spin qubits \cite{Leuenberger2001,Tejada2001,Troiani2005,Affronte2009,Ardavan2009,Stamp2009} or spin-based quantum memories because of several attractive characteristics: the ability to tune their properties, e.g. spin, magnetic anisotropy, resonance frequencies, etc, by chemical design and their high spins (e.g. $S=10$ for both Fe$_{8}$ and Mn$_{12}$), large densities (typically $\sim 10^{20}-10^{21}$ spins/cm$^{3}$), and the fact that, in many SMM crystals, the anisotropy axes of each magnetic centre are aligned parallel to each other, which might enable the attainment of stronger couplings than those previously achieved with other natural spin systems.

Here, we study from a theoretical perspective the specific case of high-spin magnetic molecules coupled to some quantum circuits, namely superconducting coplanar resonators and flux qubits. We examine, on the one hand, the possibility that strong coupling to single molecules might be achieved in the near future with available technologies and, on the other, what new physics, or new physical regimes, can be expected from the coupling of SMMs crystals to these devices. In this way, we aim to provide some guidance to future experimental work in this field.

The paper in organized as follows. Section \ref{SMMs} describes the basic features and the spin Hamiltonian of SMMs. A generic framework to calculate the magnetic coupling to electromagnetic rf fields is introduced, and then applied to discuss how such a coupling depends on molecular properties, such as spin and anisotropy, as well as on the intensity and orientation of the external magnetic field. The following two sections \ref{resonators} and \ref{fluxqubits} give realistic estimates of the coherent coupling of SMMs to superconducting coplanar resonators and flux qubits, respectively, as a function of their dimensions and geometries. The final section \ref{conclusions} gives the conclusions of the present work and discusses possible experimental implementations.

\section{Coupling of single molecule magnets to quantum radiation fields}
\label{SMMs}

\subsection{Basic properties and spin Hamiltonian of a SMM}

The magnetic configuration of a SMM is mainly determined by exchange couplings between the ions that form its magnetic core and by their interactions with the crystal field. The former give rise to multiplets with well-defined spin values, while the latter generate a magnetic anisotropy, thus also a zero-field splitting within each multiplet. Here, we consider only the ground state multiplet $S$ and neglect its quantum mixing with excited multiplets. This approximation, widely used to describe the physics of SMMs, is known as the "giant spin approximation". The effective spin Hamiltonian of a SMM reads then as follows
\begin{equation}
{\cal H}_{\rm s} = \sum_{k,l} B_{k}^{l}O_{k}^{l}-g_{S}\mu_{\rm B} \left( B_{X}S_{X} + B_{Y}S_{Y} + B_{Z}S_{Z} \right)
\label{GSHamiltonian}
\end{equation}
\noindent where $O_{k}^{l}$'s are Stevens effective spin operators \cite{Stevens1952}, $B_{k}^{l}$'s are the corresponding magnetic anisotropy parameters, $g_{S}$ is the gyromagnetic ratio and $B_{X}$, $B_{Y}$, and $B_{Z}$ are the components of an external magnetic field along the molecular axes $X$, $Y$, and $Z$. The molecular symmetry and structure determine which anisotropy parameters are nonzero as well as their relative intensities.

\begin{figure}
\includegraphics[width=\textwidth]{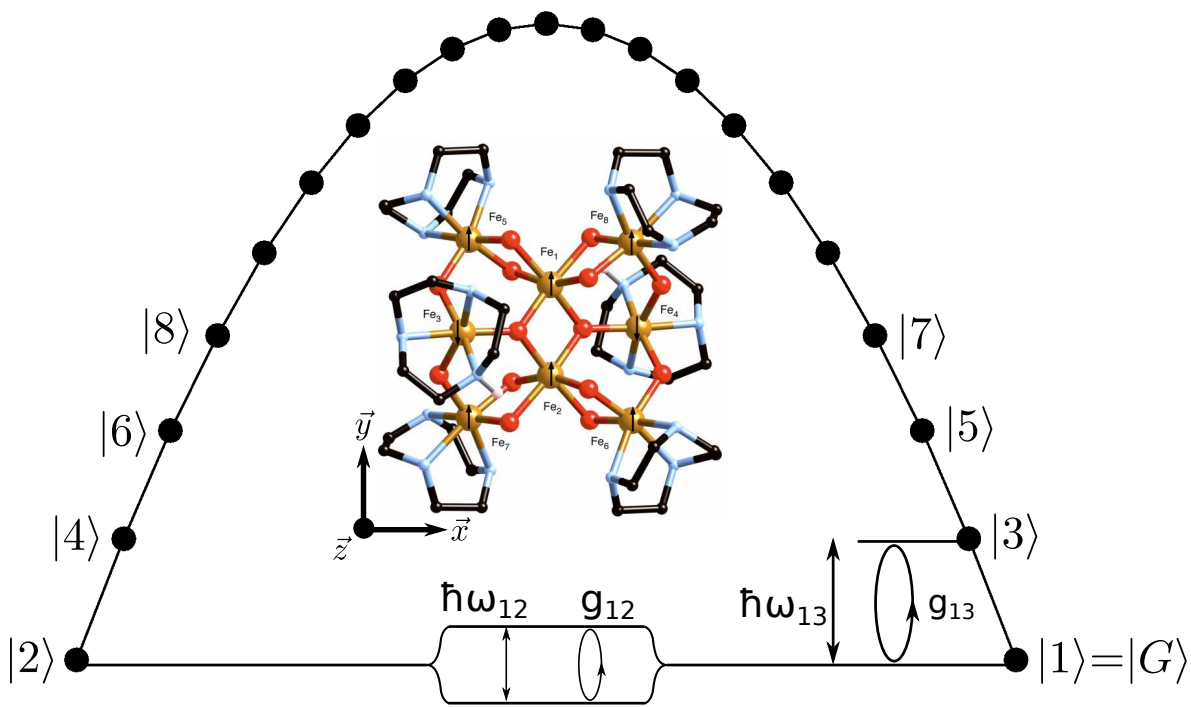}
\caption{Energy level scheme of the [(C$_{6}$H$_{15}$N$_{3}$)$_{6}$Fe$_{8}$O$_{2}$(OH)$_{12}$] single molecule magnet, shown in the inset and referred to in shorthand as Fe$_{8}$. Two possible selections of states for the use of this SMM as qubit are schematically shown.}
\label{Fig1_Fe8}
\end{figure}

One of the simplest situations corresponds to a spin with Ising-like second order anisotropy, which corresponds to $B_{2}^{0} < 0$ and all other terms being zero, i.e. to a spin Hamiltonian

\begin{equation}
{\cal H}_{\rm s} = B_{2}^{0} \left[ 3S_{Z}^{2} - S(S+1) \right] - g_{S}\mu_{\rm B} \left( H_{X}S_{X} + H_{Y}S_{Y} + H_{Z}S_{Z} \right)
\label{GSHamiltonian_Ising}
\end{equation}

\noindent As Fig. \ref{Fig1_Fe8} shows, such a "diagonal" anisotropy splits the $S$ multiplet into a series of doublets, associated with eigenstates $|\pm m \rangle$ of $S_{Z}$. As a function $m$, the energy shows then a characteristic double-well potential landscape. Off-diagonal anisotropy terms [i.e. those having $l \neq 0$ in Eq. (\ref{GSHamiltonian})] induce quantum tunneling, across the magnetic anisotropy barrier, between states $|+m \rangle$ and $|-m \rangle$ and remove their initial degeneracy by a quantum tunnel splitting $\Delta_{m}(0)$. The degeneracy can also be removed by the application of an external magnetic field $\vec{H}$. Energy splittings can be tuned, to some extent, by varying the intensity and orientation of $\vec{H}$ (see Figs. \ref{Fig2_gvsHz}, \ref{Fig3_gvsHz}, and \ref{Fig4_gvsHy}). In particular, close to $B_{Z}=0$, the splitting between the first excited and ground states $\hbar \omega_{12} \simeq \sqrt{ \left[\Delta_{S}(\vec{B}) \right]^{2} + \xi_{S}^{2} }$, where $\Delta_{S}(\vec{B})$ is the ground doublet field-dependent quantum tunnel splitting and $\xi_{S} = 2g_{S} \mu_{\rm B} B_{Z} S$ is the magnetic bias. The magnetic field enables also the initialization of the SMM state. For $S=10$, and at $T=0.1$ K, the thermal population of the ground state becomes $\gtrsim 99.99$ \% for $\mu_{0} H_{Z} \gtrsim $ 34 mT.

It is worth mentioning here that Eq. (\ref{GSHamiltonian}) applies also to, e.g., NV centres in diamond, which have $S = 1$ and a zero-field splitting determined by second-order anisotropy terms with $B_{2}^{0} \simeq 2.88$ GHz ($0.144$ K) and $B_{2}^{2}/B_{2}^{0} \lesssim 3.5 \times 10^{-3}$. Therefore, the theoretical framework that follows will enable us to compare both situations.

\subsection{Coupling of a SMM to a quantum electromagnetic radiation field}

The coupling between a spin, described by the Hamiltonian ${\cal H}_{\rm s}$, and a superconducting quantum circuit, described by ${\cal H}_{\rm q}$, is governed by the Zeeman interaction,
\begin{equation}
{\cal H} = {\cal H}_{\rm q} + {\cal H}_{\rm s} - \left( \vec{W}^{{\rm (q)}} V_{\rm q}\right)\vec{S}
\label{eq:zeemancoup}
\end{equation}
\noindent where $\vec{W}^{\rm (q)} = g_{S}\mu_{B}\vec{B}^{\rm (q)}$ is proportional to the magnetic field $\vec{B}^{\rm (q)}$ generated by the superconducting circuit at the spin position and $V_{\rm q}$ is an operator acting on the circuit's variables.

For the present purposes, the spin can be treated as a two-level system. This is possible by focusing only on those two spin levels whose energy difference is in (near) resonance with the circuit's transition frequency $\hbar \omega$. More specifically, we choose the spin ground state $|G\rangle$ and one excited state $|E \rangle$. Two possible choices, relevant to real SMMs, are shown in Fig. \ref{Fig1_Fe8}. We define the spin transition frequency
\begin{equation}
\hbar\omega_{\rm{G,E}} \equiv \langle E|{\cal H}_{\rm s}|E \rangle - \langle G|{\cal H}_{\rm s}|G\rangle
\label{resonantfrequency}
\end{equation}
\noindent and the transition matrix element
\begin{eqnarray}
\hbar g &\equiv& \langle G|\vec{W}^{(q)}\vec{S} |E \rangle \\ \nonumber
&=& W^{(q)}_{X} \langle G|S_{X} |E \rangle + W^{(q)}_{Y} \langle G|S_{Y} |E \rangle+ W^{(q)}_{Z} \langle G|S_{Z} |E \rangle
\label{Melements}
\end{eqnarray}
\noindent Achieving strong coupling requires that the SMMs can be tuned to resonance with the circuit, i.e. that $\hbar \omega_{\rm{G,E}} \simeq \hbar\omega$ for a given $|E\rangle$, and that the relevant matrix element of the Zeeman interaction is sufficiently large. In the remainder of this section, we discuss how matrix elements $\langle G| S_{I} |E \rangle$, with $I = X, Y, Z$, thus also $g$, depend on the choice of state $|E \rangle$ as well as on the magnetic anisotropies and experimental conditions that can be met with real SMMs. The actual coupling $g$ depends also on the magnetic field generated by a given circuit, thus on its design and geometry. These aspects will be considered in sections \ref{resonators} and \ref{fluxqubits} below.

\subsection{Calculation of transition matrix elements}

\subsubsection{Transitions between zero-field split levels}

\begin{figure}
\includegraphics[width=\textwidth]{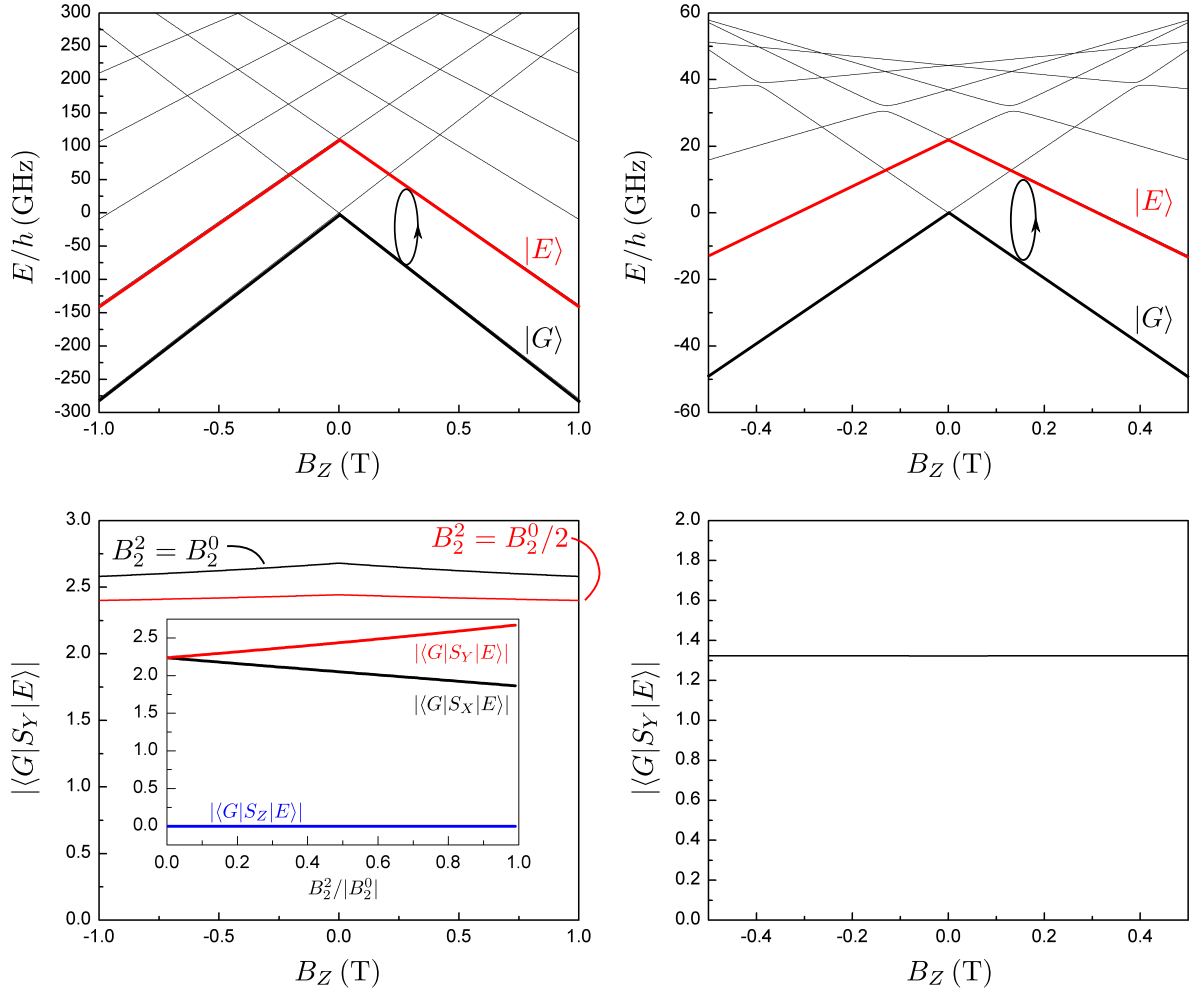}
\caption{Top: energy levels of a generic SMM ($S=10$, $B_{2}^{0}/k_{\rm B} = -0.1$ K, and $B^{2}_{2}/k_{\rm B} = 0.1$ K) on the left and of GdW$_{10}$ ($S=7/2$, $B_{2}^{0}/k_{\rm B} = -0.059$ K, and $B^{4}_{4}/k_{\rm B} = 4 \times 10^{-4}$ K) on the right as a function of the external field parallel to the easy axis ($B_{Z}$). Bottom: spin matrix elements associated with transitions between zero-field split levels (i.e., between states $|G\rangle = |1 \rangle$ and $|E\rangle = |3 \rangle$ of Fig. \ref{Fig1_Fe8}). For the generic SMM, matrix elements of two different values of $B_2^{2}$ are shown. The levels associated with the computational basis $|G\rangle$ and $|E\rangle$ are marked with thicker lines and labelled in the energy level diagrams. The inset shows the transition matrix elements for all three components of $\vec{S}$ as a function $B_2^{2}$.}
\label{Fig2_gvsHz}
\end{figure}

In this and the next subsections, we consider a generic $S=10$ SMM with ${\cal H}_{\rm s}$ described by Eq. (\ref{GSHamiltonian}) and second order anisotropy terms only. This situation applies to some of the best known SMMs, such as Fe$_{8}$, shown in the inset of Fig. \ref{Fig1_Fe8}, or even Mn$_{12}$. A first choice, reminiscent of the situation met with NV centres in diamond, is to identify $|E \rangle$ with state $|3 \rangle$, as shown in Fig. \ref{Fig1_Fe8}. Neglecting $B_{2}^{2}$, which plays a minor role here unless it is comparable to $B_{2}^{0}$ and $B_{Z} \simeq 0$ (see Fig. \ref{Fig2_gvsHz}), this situation corresponds to $|G \rangle \simeq |+S \rangle$ and $|E \rangle \simeq |+S-1 \rangle$ for $B_{2}^{0} < 0$ and to $|G \rangle \simeq |0 \rangle$ and $|E \rangle \simeq |+1 \rangle$ for $B_{2}^{0} >0$. The splitting $\hbar \omega_{12} = \hbar \omega_{12}(0) + g\mu_{\rm B}B_{Z}$, where $\hbar \omega_{12}(0) = 3(2S-1)B_{2}^{0}$ in the former case and $\hbar \omega_{12}(0) = 3B_{2}^{0}$ in the latter case. Relevant transition matrix elements correspond to transverse spin components $S_{X}$ and $S_{Y}$ and can be calculated analytically

\begin{equation}
g \propto \frac{1}{2} \sqrt{\left(S-m_{\rm gs}\right)\left( S + m_{\rm gs} + 1 \right)}
\label{g13_for easy spin}
\end{equation}
\noindent where $m_{\rm gs}$ is the $S_{z}$ eigenvalue of the ground state. Depending on the sign of the anisotropy, Eq. (\ref{g13_for easy spin}) gives
\begin{eqnarray*}
B_{2}^{0} > 0 & \Rightarrow & m_{\rm gs} = 0 \Rightarrow g \propto \frac{1}{2}\sqrt{S (S+1)}\\
B_{2}^{0} < 0  & \Rightarrow & m_{\rm gs} = +S \Rightarrow g \propto \frac{1}{2}\sqrt{2S}
\label{g13_for easy spin2}
\end{eqnarray*}
\noindent High couplings are therefore achieved for high-spin $S$ materials, optimally with $B_{2}^{0} > 0$. Furthermore, $g$ is but weakly affected by external magnetic fields (see Fig. \ref{Fig2_gvsHz}). A difficulty associated with this choice of basis is that the zero-field splittings of high-spin SMMs, such as Fe$_{8}$ or Mn$_{12}$, are often very large (e.g. $\hbar \omega_{12} (0) \simeq 114$ GHz for Fe$_{8}$) as compared with the typical resonance frequencies of either superconducting resonators \cite{Blais2004,Wallraff2004} ($\omega/2\pi \simeq 1$ to $40$ GHz) or gap-tunable flux qubits \cite{Zhu2010} (for which $\omega/2\pi \simeq 1-10$ GHz).

\subsubsection{Transitions between 'spin-up' and 'spin-down' states: photon induced quantum tunneling}

\begin{figure}
\includegraphics[width=\textwidth]{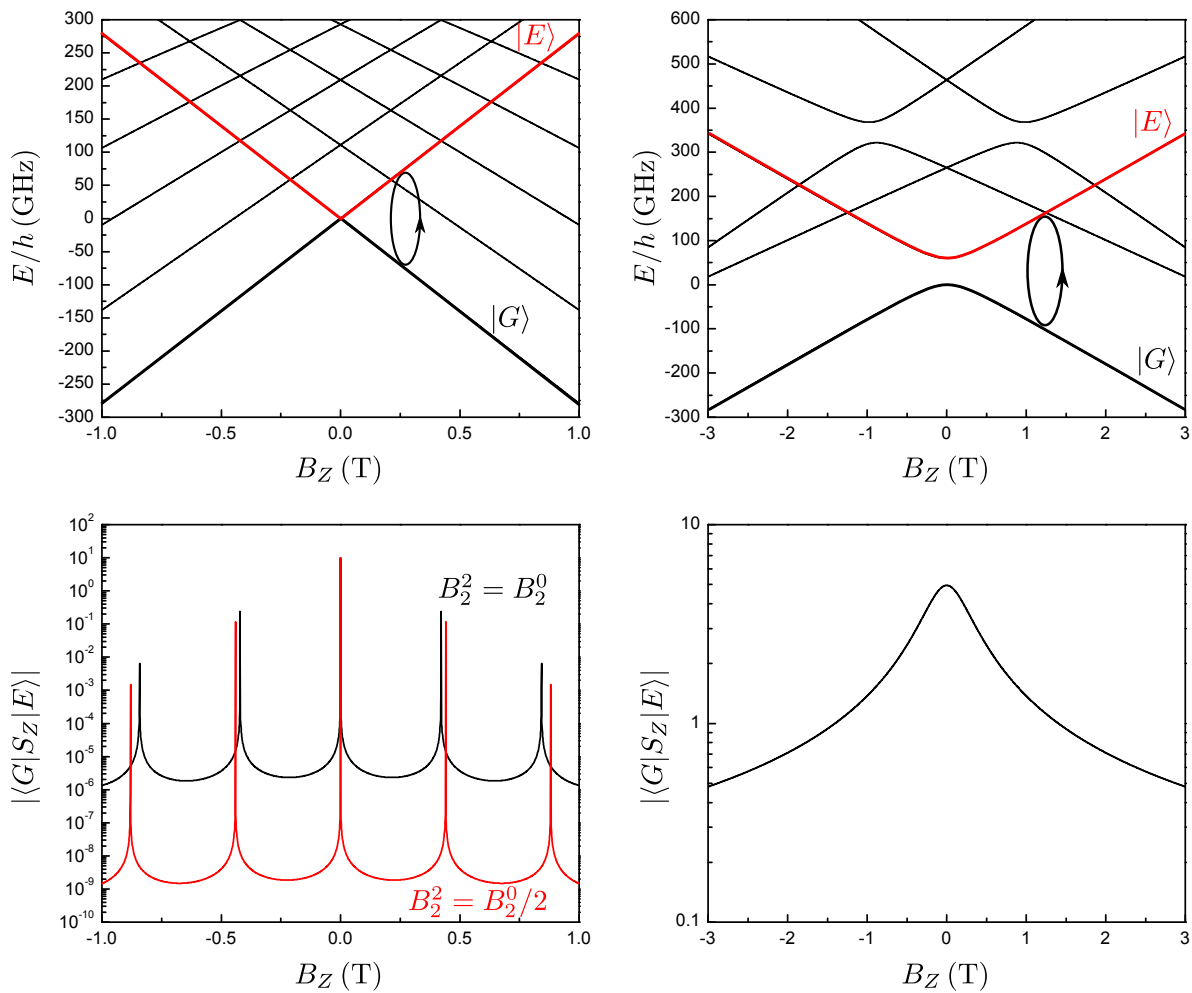}
\caption{Top: energy levels of a generic SMM (same parameters as those used in Fig. \ref{Fig2_gvsHz}) on the left and of TbW$_{30}$ ($S=6$, $B_{2}^{0}/k_{\rm B} = -1.0$ K, $B_{4}^{0}/k_{\rm B} = 6 \times 10^{-3}$ K, $B_{6}^{0}/k_{\rm B} = -1.1 \times 10^{-5}$ K, $B_{6}^{5}/k_{\rm B} = 1.7 \times 10^{-3}$ K) on the right as a function of the external field parallel to the easy axis ($B_Z$). Bottom: spin matrix elements associated with transitions between states of the ground state doublet (i.e., between states $|G\rangle = |1 \rangle$ and $|E\rangle = |2 \rangle$ of Fig. \ref{Fig1_Fe8}). For the generic SMM case, the matrix elements of two different values of $B_2^{2}$ are shown. The two levels associated with the computational basis $|G\rangle$ and $|E\rangle$ are marked with thicker lines and labelled in the energy level diagrams.}
\label{Fig3_gvsHz}
\end{figure}

A second natural choice is to use, as "computational" basis for the spin qubit, the two lowest-lying eigenstates of ${\cal H}_{\rm s}$ at zero field, which we denote here (see Fig. \ref{Fig1_Fe8}) by $|1 \rangle$ and $|2 \rangle$. For $B_{2}^{2} = 0$, these states correspond to degenerate 'up' and 'down' spin orientations, thus all matrix elements vanish. Off-diagonal anisotropy terms give rise to a finite $\hbar \omega_{12} = \Delta_{S}(0)$. At zero-field, $|G \rangle \simeq (1/\sqrt{2}) ( |+S \rangle + |-S \rangle )$ and $|E \rangle \simeq (1/\sqrt{2}) ( |+S \rangle - |-S \rangle )$, thus $\langle G|S_{Z} |E \rangle \simeq S$, the other elements being close to zero. This is confirmed by numerical results shown in Fig. \ref{Fig3_gvsHz}. Considering the high spin of SMMs, this transition can therefore give rise to potentially strong couplings. However, $\Delta_{S}(0)$ often lies in the region of micro-Kelvins or even smaller. For instance, $\Delta_{S} \simeq 10^{-7}$ K ($10^{-11}$ K) or barely $2.1$ kHz ($0.2$ Hz) for Fe$_{8}$ (Mn$_{12}$). A magnetic field needs then to be applied in order to tune $\omega_{12}$ to the circuit frequencies.

Maximum energy changes are obtained when $\vec{B}$ is oriented along the easy magnetization axis $Z$ (Fig. \ref{Fig3_gvsHz}). However, any bias $\xi_{S} \gtrsim \Delta_{S}(0)$ effectively suppresses the overlap between the wavefunctions of $|1 \rangle$ and $|2 \rangle$ states (that effectively become $|+S \rangle$ and $|-S \rangle$ states) resulting in a dramatic decrease of $g$ with increasing $B_{Z}$. The matrix elements show, in fact, narrow peaks at those values of $B_{Z}$ that induce level anti-crossings. These resonances are associated with a photon induced tunneling process between quasi-degenerate spin states. Resonances occur only at every even numbered level crossings (i.e. for $B_{Z} \simeq nB_{1}$, with $B_{1} = 3B_{2}^{0}/g_{S}\mu_{\rm B}$ and $n = 0, 2, \ldots$) because $B_{2}^{2}O_{2}^{2}$ only mixes states $|m \rangle$ and $|m^{\prime} \rangle$ such that $m-m^{\prime}$ is even. The width of each resonance (thus also the field region of potential interest for coupling to a circuit) can be increased by enhancing the off-diagonal parameter $B_{2}^{2}$, although it nevertheless remains very narrow even for the maximum $B_{2}^{2} = B_{2}^{0}$.

\begin{figure}
\includegraphics[width=\textwidth]{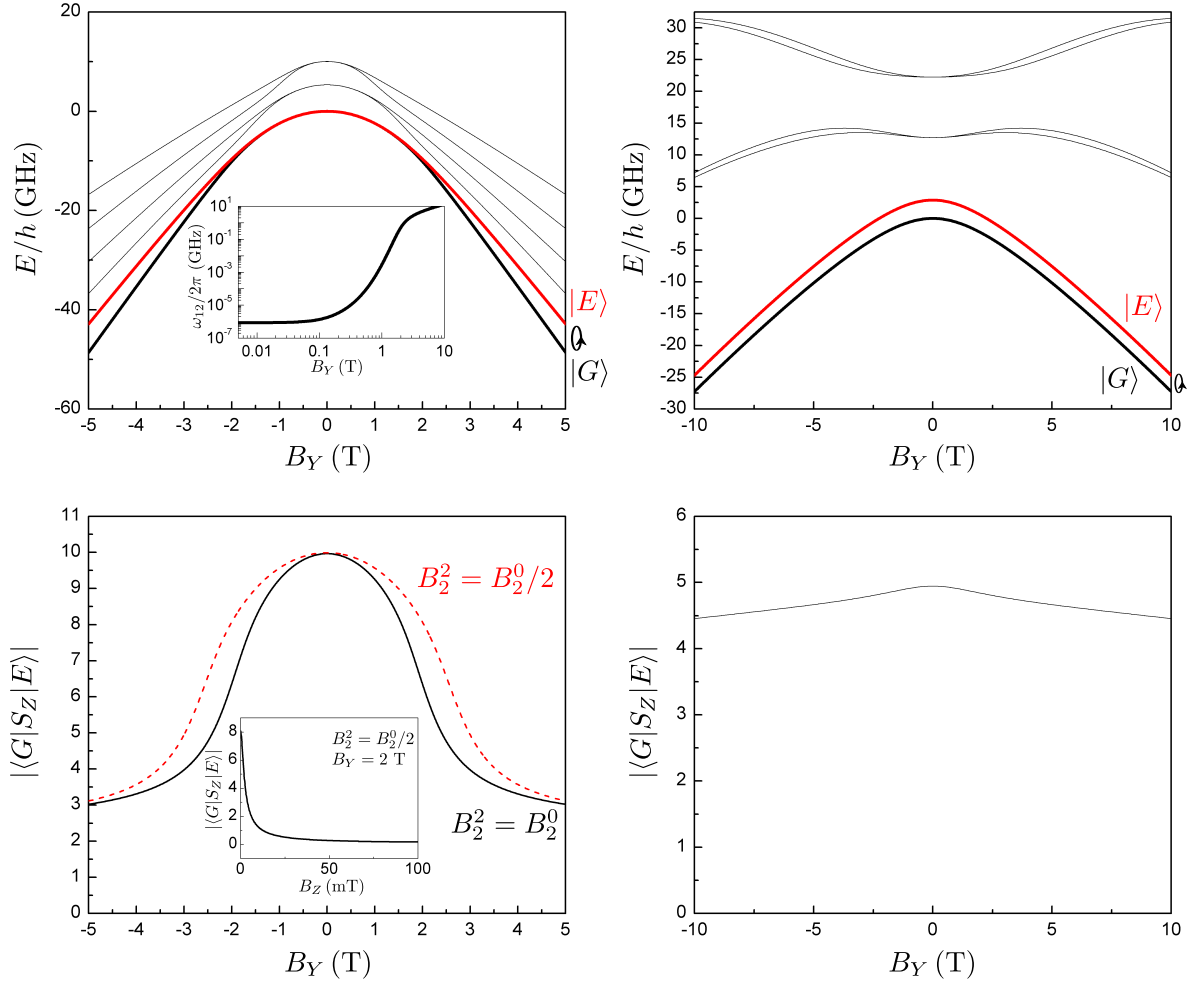}
\caption{Top: energy levels of a generic SMM on the left and of TbW$_{30}$ (same parameters as those used in Fig. \ref{Fig3_gvsHz}) on the right as a function of the external field parallel to the medium anisotropy axis ($B_Y$). The inset shows the energy difference, in GHz, between $|E\rangle$ and $|G\rangle$ also as a function of $B_Y$. Bottom: spin matrix element associated with transitions between states of the ground state doublet (i.e., between states $|G\rangle = |1 \rangle$ and $|E\rangle = |2 \rangle$ of Fig. \ref{Fig1_Fe8}). For the generic SMM case, the matrix elements of two different values of $B_2^{2}$ are shown. The two levels associated with the computational basis $|G\rangle$ and $|E\rangle$ are marked with thicker lines and labelled in the energy level diagrams. The bottom inset shows the dependence of the spin matrix element on $B_{Z}$ for $B_{Y} = 2$ T.}
\label{Fig4_gvsHy}
\end{figure}

Alternatively, $\hbar \omega_{12}$ can also be tuned, while retaining a strong overlap between $|1 \rangle$ and $|2 \rangle$, thus a high $\langle G|S_{Z} |E \rangle$ (see Fig. \ref{Fig4_gvsHy}), by a transverse magnetic field $B_{Y}$. This is a highly nonlinear effect (see the inset of Fig. \ref{Fig4_gvsHy}), meaning that strong magnetic fields are required to make $\omega_{12}$ close to $\omega$.  The use of stronger magnetic fields also imposes stringent conditions to the alignment of $\vec{B}$ which, as follows from the data shown in the bottom inset of Fig. \ref{Fig4_gvsHy}, cannot deviate more than about $0.5$ deg. from the $XY$ plane.

\subsubsection{Single ion magnets vs single molecule magnets}

The previous results show that, because of their high spin values, the coupling of SMMs to a rf magnetic field can attain very high values. However, the also high magnetic anisotropy barriers (they tend to increase with $S$) and correspondingly small quantum tunnel splittings can pose some important technical difficulties: the use of very high frequencies to attain resonant conditions with zero-field split levels ($\gtrsim 110$ GHz for Fe$_{8}$ or $\gtrsim 220$ GHz for Mn$_{12}$) or the need of applying strong and very accurately aligned magnetic fields if one focuses on transitions within the tunnel split ground state doublet. In addition, achieving a pre-designed control over relevant parameters (spin and magnetic anisotropies) is a difficult task, if feasible at all, with polynuclear clusters.

Mononuclear SMMs (or single ion magnets SIMs \cite{Ishikawa2005,AlDamen2008,Aldamen2009}) are, by contrast, much simpler: they consist of just one magnetic ion, often a lanthanide, encapsulated inside a non magnetic shell of ligand molecules. These materials can be seen as the molecular analogues to diluted lanthanide salts, which are also seen as promising spin qubits \cite{Bertaina2007}. An advantage of molecular SIMs over these materials is that the local coordination of the magnetic ion can be modified by adequately choosing the nature and structure of the ligand shell, thus providing a rich playground for the rational design of their spin Hamiltonian \cite{Martinez-Perez2012}.

\begin{figure}
\centering
\includegraphics[width=0.5\textwidth]{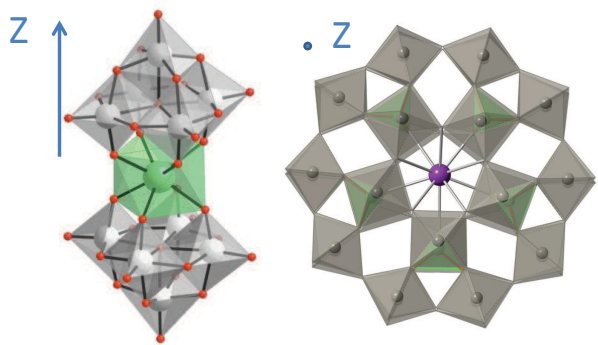}
\caption{Molecular structures of LnW$_{10}$ (left) and of LnW$_{30}$ (right) single ion magnets.}
\label{Fig5_SIMs}
\end{figure}

Some specific examples can help to understand how the problems mentioned above can be overcome with the use of simpler molecules. Here, we discuss some possibilities offered by two families of SIMs based on polyoxometalate complexes (see Fig. \ref{Fig5_SIMs}). If one seeks to reduce the magnetic anisotropy, thus also the zero-field splitting $\hbar \omega_{13}$, the use of Gd$^{3+}$ ions is a good option because of their close to spherical electronic configuration. In addition, the sign and intensity of the magnetic anisotropy are determined, to a large extent, by the local coordination \cite{Martinez-Perez2012}. In the elongated GdW$_{10}$ molecule, for instance, the preferred magnetization axis points along the molecular axis $Z$, giving rise to a ground state $m_{\rm gs} \pm 7/2$ separated from the first excited doublet $\pm 5/2$ by a small zero-field splitting. This leads to a rather convenient value for the resonance frequency $\omega_{13}/2 \pi \simeq 20$ GHz. For the donught-shaped GdW$_{30}$ molecule, the anisotropy is even weaker and of opposite sign (i.e. $B_{2}^{0} > 0$), thus $Z$ becomes a hard magnetization axis and the easy axis ($Y$), lying within the molecular plane, is determined by the presence of strong off-diagonal anisotropy terms. At zero field, the splitting is $\hbar \omega_{13} \simeq 6.4$ GHz, thus in tune not only with coplanar resonators but also with gap-tunable flux qubits. In fact, all spin levels of GdW$_{30}$ lie within a frequency band of about $20$ GHz, accessible to most superconducting circuits.

For lanthanide ions other than Gd$^{3+}$, the magnetic anisotropy and the ground state depend also on intrinsic electronic structure of the ion itself. A particularly interesting situation is found for TbW$_{30}$. The combination of second- fourth- and sixth-order diagonal anisotropy terms gives rise to a ground state doublet with $m_{\rm gs} = \pm 5$ \cite{Cardona-Serra2012}. More importantly, the fivefold molecular symmetry allows the presence of a strong $B_{6}^{5} O_{6}^{5}$ term, which efficiently mixes these spin states. The result is a two-level spin system characterized by a large zero-field quantum tunnel splitting $\Delta_{S}(0) \sim 60$ GHz and therefore a high transition matrix element $g$. More importantly, $g$ is very robust against the action of external magnetic fields (see Figs. \ref{Fig3_gvsHz} and \ref{Fig4_gvsHy}).

\section{Coupling of SMMs to superconducting coplanar resonators}
\label{resonators}

\subsection{Device description and parameters}
\begin{figure}
\centering
\includegraphics[width=0.65\textwidth]{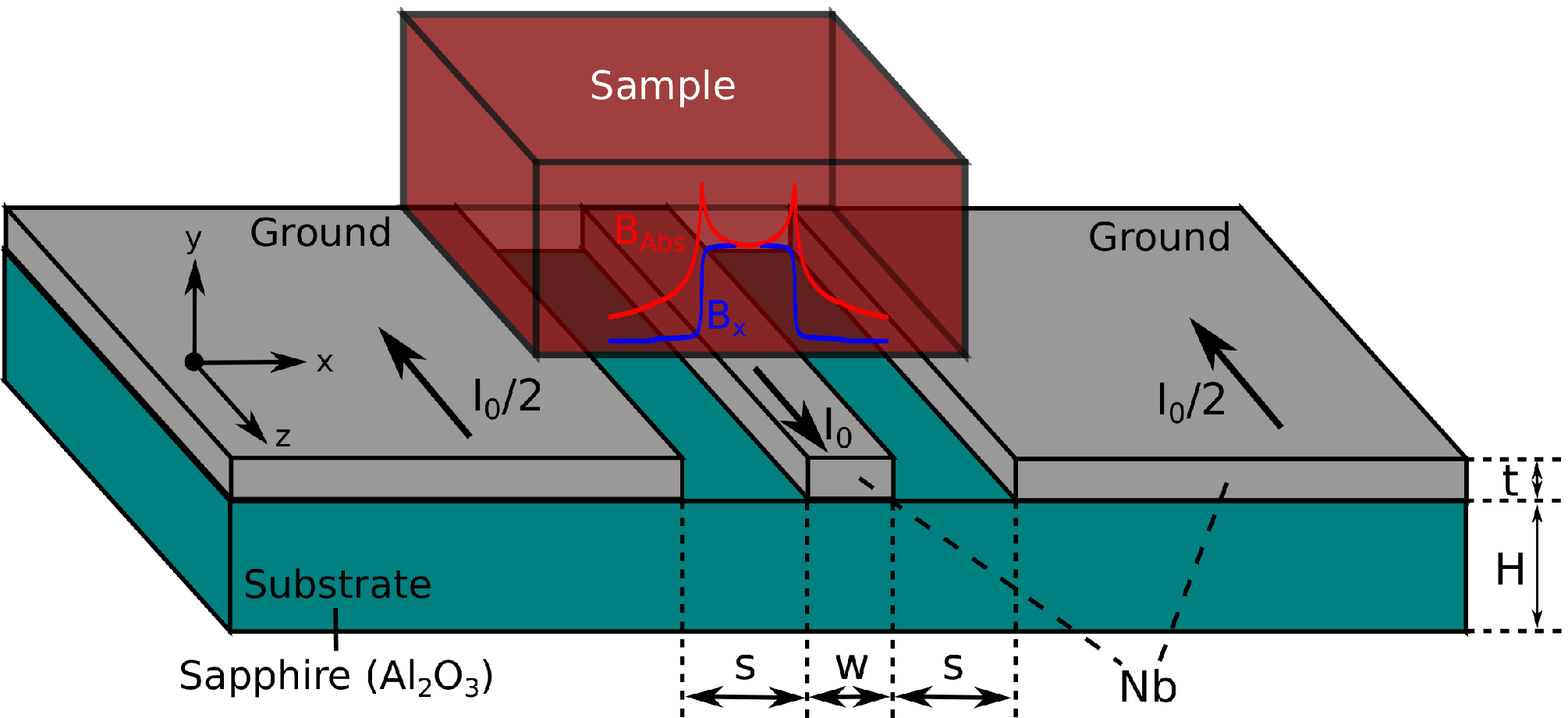}\\[5mm]
\begin{tabular}{|c|c|}
\hline
\multicolumn{2}{|c|}{Resonator dimensions} \\
\hline
s & 7 $\mu$m \\
\hline
w & 50 nm to 14 $\mu$m \\
\hline
t & 150 nm \\
\hline
H & 75 $\mu$m \\
\hline
\end{tabular}
\begin{tabular}{|c|c|}
\hline
\multicolumn{2}{|c|}{Sample dimensions} \\
\hline
Width & 40 $\mu$m \\
\hline
Length & 40 $\mu$m \\
\hline
Height & 0.1 to 75 $\mu$m \\
\hline
\multicolumn{2}{|c|}{} \\
\hline
\end{tabular}
\caption{Basic geometry and dimensions of the CPW resonator and the magnetic samples used for the calculations described in this work}
\label{CPWG_diagram}
\end{figure}

Coplanar resonators are microwave devices that consist of a $\lambda/2$ section of a coplanar waveguide (CPW) that is coupled to external feed lines via gap capacitors. A schematic diagram of such a device is shown in Fig. \ref{CPWG_diagram}. The fundamental mode resonant frequency is determined by the length of the resonator through the equation $f_0 = \frac{c}{\sqrt{\epsilon_{\rm eff}}}\frac{1}{2l}$. Here $\epsilon_{\rm eff}$ is the effective dielectric constant of the CPW and depends on the waveguide geometry and the dielectric constants of the surrounding media \cite{Simons2004}. As with transmission lines, the electromagnetic mode is described as a voltage and current wave where the current in the centre line is equal and opposite to the current in the ground plates.

Making the resonator out of superconducting materials, such as Nb or NbTi, and using low loss dielectric substrates, such as sapphire, helps to reduce the losses in the system and allows the reduction of the resonator cross section down to the micrometer level while maintaining quality factors of up to $10^{5}-10^{6}$ \cite{Frunzio2005,Barends2007,Goppl2008}.

\subsection{Coherent coupling to individual SMMs and SMM ensembles}

For a resonator, Eq. (\ref{eq:zeemancoup}) takes the following form \cite{Verdu2009}:
\begin{eqnarray}
\nonumber
{\cal H} & = & {\cal H}_{\rm s} + {\cal H}_{\rm res} - \vec{\mu}\cdot \vec{B}^{\rm (q)} \\
& = & {\cal H}_{\rm s} +
\hbar a^\dagger a +
\hbar  g(\vec r_j) \sigma_x (a  -a^\dagger) \label{eq:8}
\end{eqnarray}
\noindent
where we have projected onto the basis formed by the two relevant SMM states $|G\rangle$ and $|E\rangle$,  $\sigma_x$ is the Pauli matrix acting on this basis and the coupling strength:
\begin{equation}
g(\vec r_j) = g_S \mu_{\rm B}\left| \langle {\rm G}| \vec{b}_{\rm rms} (\vec r_j)\vec{S} | {\rm E}\rangle \right|
\label{interaction}
\end{equation}
with $ \vec{b}_{\rm rms} (\vec r_j)$ the root mean square value of the field $\vec{b}$ generated by the vacuum current (see below).  The position $\vec r_j$ matches the spin location.  Through this section we will assume that the magnetic sample is centered at the maximum magnetic field strength generated by the resonator, i.e. at the midpoint of the resonator.

The coupling per spin is usually small (of the order of a few 100 Hz, see below) and losses can easily overcome the coherent coupling.  Therefore, we will also consider the coupling to an ensemble, e.g. a crystal, of $N$ SMMs. For this, we sum (\ref{interaction}) over each spin at position $\vec r_{\rm j}$.  It is convenient to introduce the collective spin operator
\begin{equation}
\label{collective}
b^{\dagger}=\frac{1}{\sqrt{N}\bar{g}}\sum_{j}^{N}g_{j}^{*}\sigma_{j}^{+}
\end{equation}
where $\bar{g}$ is the average coupling, defined as $\bar{g}^{2} \equiv \sum_{j} \left| g_{j} \right|^{2}/N$. In the low polarization level $\langle \sum \sigma_j^\dagger \sigma^-_j \rangle \ll N$ these operators approximately fulfill bosonic commutation relations, $[b,b^{\dagger}] \approx 1$ \cite{Hummer2012}. Equation (\ref{eq:8}) then becomes approximately equal to the Hamiltonian of two coupled resonators,
\begin{equation}
{\cal H}  =  {\cal H}_{\rm q} + {\cal H}_{\rm s} - \hbar g_{N} (b^{\dagger} + b) (a^\dagger + a)
\label{eq:collective}
\end{equation}
with an effective coupling given by,
\begin{equation}
g = g_s\frac{\mu_B}{h}\sqrt{n\int_V \left| \langle G|\vec{b}_{\rm rms}\cdot\vec{S}|E\rangle \right|^2dV} \label{coupling}
\end{equation}
where we have replaced the sums by integrals and assumed a uniform density $n$. Let us emphasize that Eq. (\ref{coupling}) leads to a $\sqrt{N}$ enhancement of the effective coupling with respect to that of a single spin.

In order to calculate the collective coupling of SMMs to a single photon, one needs to evaluate the magnetic field generated by the rms of the vacuum current fluctuations, $I_{\rm rms}$.  This current can be found considering that the zero point energy of the resonator is shared equally between the electric and magnetic fields:
\begin{equation}
\frac{\hbar \omega}{4} = \frac{1}{2}LI^{2}_{\rm rms} \quad \Rightarrow \quad
I_{\rm rms}=\omega \sqrt{\frac{\hbar\pi}{4Z_0}}
\end{equation}
where $L=2Z_0/(\pi\omega)$ is the lumped inductance of the resonator \cite{Pozar2011} and $Z_0$ is the characteristic impedance of the transmission line segment that forms the resonator.  Taking a standard value of $Z_0\simeq 50\,\Omega$, we find that the vacuum current fluctuations are of about $8$ nA/GHz. For $\omega/2 \pi \lesssim 37$ GHz, the current in a resonator is therefore somewhat smaller than the current $I_{p} \sim 0.3 \mu$A circulating via a flux qubit close to its compensation point, so we can then expect smaller couplings.

We use the Comsol Multiphysics AC/DC module to calculate the field distribution given the resonator geometry and currents.  We model only a cross section of the resonator so the calculated fields are approximated by those generated by an infinite conductor length. This approximation holds as long as the SMM crystals are placed close to the resonator centre and the crystal length is much shorter than that of the resonator itself, which ranges from $0.5$ to $10$ mm for the frequencies of relevance here.

Even for DC currents, the current density distribution in a superconductor is not uniform and different from that of a normal conductor. In real superconductors, the superconducting current density decays exponentially with the distance to its surface and the decay constant is the London penetration depth $\lambda_{\rm L}$. For Nb, $\lambda_{\rm L} \simeq 80$ nm at $4$ K and increases as temperature increases toward the critical temperature ($T_{\rm c} \simeq 9$ K) \cite{Kim2003}. Since the thickness of the superconducting lines we consider (see figure \ref{CPWG_diagram}) is of this order or smaller, we need to simulate the current distribution carefully to take this effect into account. As a first approximation, we use the skin effect of standard conductors to produce the current profiles in the superconducting regions, i.e., we use alternating currents and tune the frequency $\omega_{\rm ac}$ and material parameters (conductivity $\sigma$ and magnetic permeability $\mu$) in the simulation to make the skin depth of the conductor
\begin{equation}
\lambda_{\rm skin} = \sqrt{\frac{2}{\sigma\omega_{\rm ac}\mu}}
\label{skineffect}
\end{equation}
equal to $\lambda_{\rm L} = 80$ nm.

\begin{figure}
\centering
\includegraphics[width=0.65\textwidth]{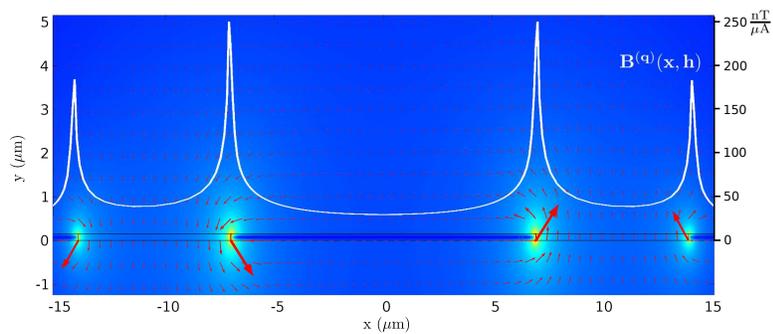}
\caption{Simulated field distribution on a CPW cross-section.  The white profile is the field value calculated at a constant distance from the substrate $y=t$, that is, right at the surface of the superconducting regions.}
\label{Bprofile}
\end{figure}

Taking all this into account, we simulate the magnetic field distribution for the geometry shown in Fig. \ref{CPWG_diagram}.  A typical magnetic field distribution is shown in Fig. \ref{Bprofile}. As expected, the superconducting current and magnetic field concentrate near the edges of the centre line and the inner edges of the ground planes. Using these magnetic field distributions and the matrix element values calculated for each SMM sample, it is possible to obtain the coupling strength from Eq. (\ref{coupling}) for crystals of varying dimensions. An appealing aspect of many SMMs crystals (including those considered here) is that the magnetic anisotropy axes of all molecules are aligned with respect to each other. This enables orienting them so that the fields from the resonator can induced the desired transitions.  Each sample and each choice of computational basis has a different optimal orientation of the magnetic anisotropy axes ($X,Y,Z$) with respect to the resonator coordinate system ($x,y,z$).  In our simulations, the axis with the largest absolute value of the transition matrix element points along the $x$-axis of the resonator (i.e. horizontal, see Figs. \ref{CPWG_diagram} and \ref{Bprofile}) while the second largest  is placed along the $y$-axis (i.e. perpendicular to the resonator).  This is because the integral of $b_X^{2}$ entering in Eq. (\ref{coupling}) is slightly larger than the $b_Y^{2}$ integral, thus leading to also slightly larger $g_N$.  We also calculate the collective coupling of NV centres in diamond crystals. In this case, one has to average over the four different orientations of their magnetic anisotropy axes. In all these calculations, we consider crystals of fixed length and width (both equal to $40 \,\mu$m, see Fig. \ref{CPWG_diagram}) and study how $g_{N}$ depends on the crystal thickness (thus also the number of spins), from $100$ nm up to $75\,\mu$m.

\begin{figure}
\centering
\includegraphics[width=\textwidth]{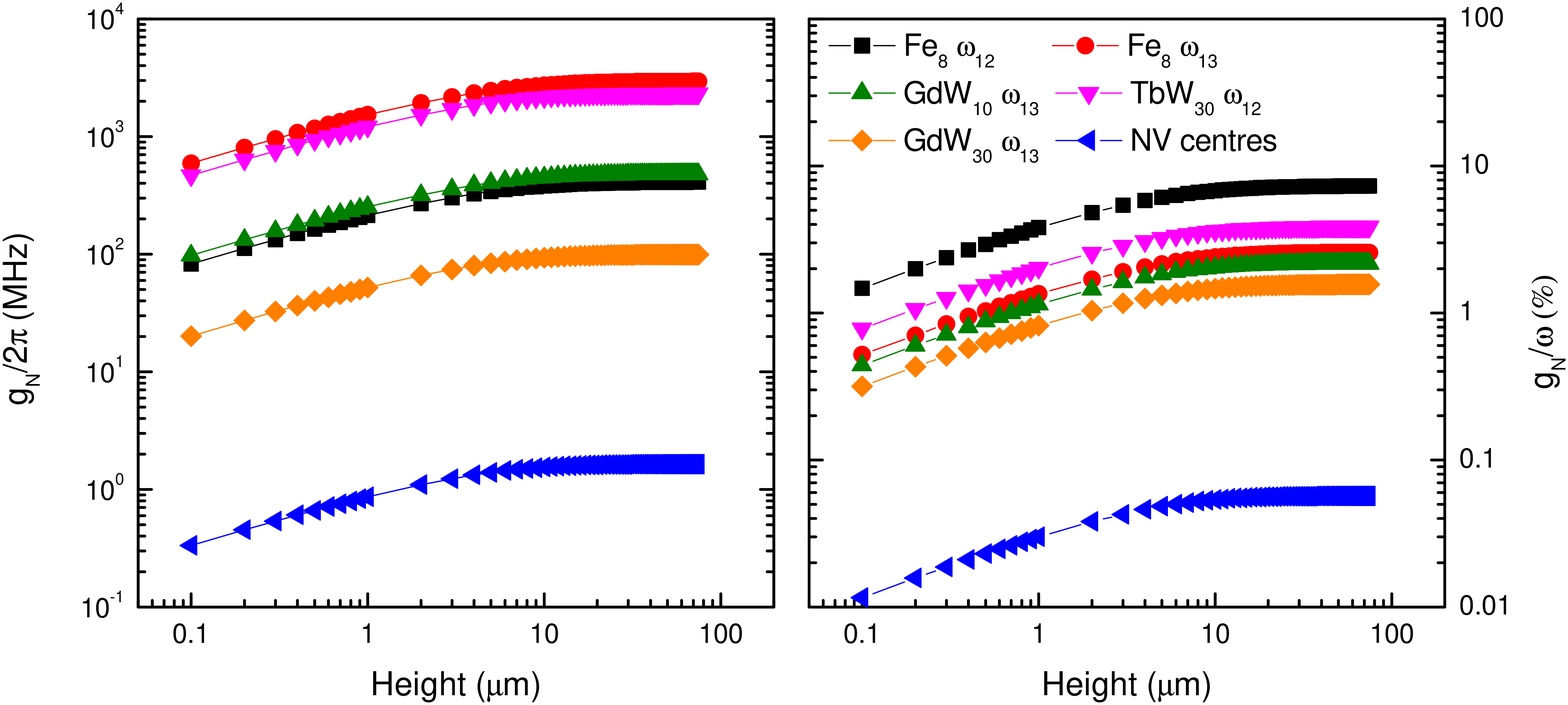}\\[5mm]
\begin{tabular}{|c|c|c|c|c|c|}
\hline
Fe$_8\;\omega_{12}$ & Fe$_8\;\omega_{13}$ & GdW$_{10}\;\omega_{13}$
& TbW$_{30}\;\omega_{12}$ & GdW$_{30}\;\omega_{13}$ & NV centres \\
\hline
5.6 GHz & 114.6 GHz & 22.1 GHz & 60 GHz & 6.4 GHz & 2.88 GHz \\
($B_Y = 2.325$ T) & & & & & \\
\hline
\end{tabular}
\caption{Coupling of $40\,\mu$m$\times 40\,\mu$m$\times {\rm thickness}$ SMM and diamond crystals to a CPW resonator as a function of crystal thickness. On the left we show the total coupling strength and on the right we show the coupling strength normalized by the resonator frequency.  For each sample, $\omega_{ij}$ denotes the transition used and the operating frequencies are detailed in the table above.}
\label{gvsh}
\end{figure}

The results are shown in Fig. \ref{gvsh} for several SMMs and different choices of their computational basis $|G \rangle$ and $|E \rangle$ as well as for NV centres in diamond. We see that the coupling first increases with crystal thickness and then saturates once the crystal is thicker than about $10-15$ $\mu$m. This behaviour reflects the decay of $\vec{b}$ with the distance $y$ from the resonator surface. It shows that only a very thin layer of spins significantly contributes to $g_{N}$ and emphasizes the importance of carefully placing the sample on top of the device. As would be expected, the dependence on the crystal thickness is essentially the same for all samples.

It can be seen that, because of their specific characteristics, the coupling to SMM crystals can be very large, much larger indeed than the coupling to NV centres in diamond crystals of equivalent size. The largest couplings $g_{N} \simeq 2-3$ GHz are found for transitions between tunnel split states of Tb$_{30}$ and between states $1$ and $3$ of Fe$_{8}$. Yet, these transitions are characterized by very high resonance frequencies $\omega_{12} \sim 60$ GHz and $\omega_{13} \simeq 114$ GHz, respectively. Very large couplings ($g_N \simeq 0.5$ GHz) are also found for transitions between tunnel split states of e.g. Fe$_{8}$, for which  $\omega_{12}$ can be tuned by applying external magnetic fields (see Fig. \ref{Fig4_gvsHy}). However, one then has to deal with rather strong ($\gtrsim 2$ T) and very accurately aligned (typically within less than $0.5$ deg.) magnetic fields (see Fig. \ref{Fig4_gvsHy}). For this reason, it might be experimentally simpler to work with Gd-based SIMs, for which $\omega_{13}$ lies between $6.4$ and $20$ GHz, and whose transition matrix elements are more robust against the action of external magnetic fields.

The couplings need to be compared with spin decoherence frequencies $\sim 1/T_{2}$, where $T_{2}$ is the phase coherence time. Experiments performed on crystals of GdW$_{10}$ and GdW$_{30}$ \cite{Martinez-Perez2012}, and of Fe$_{8}$ \cite{Takahashi2011} show that $T_{2} \lesssim 500$ ns at liquid helium temperatures and under the best conditions, thus much shorter than $T_{2} \sim 1-2$ ms of NV centres \cite{Balasubramanian2009} at room temperature. Still, the strong coupling limit $g_{N}T_{2}/2\pi \gg 1$ should be relatively easy to achieve for all these molecular materials. Furthermore, for some of the examples given in Fig. \ref{gvsh}, $g_{N}$ can in fact become a sizeable fraction of the resonator frequency, thus opening the possibility to reach and explore the ultra-strong coupling limit with a spin ensemble.

\subsection{Nanoscale resonators}

The simulations described in the previous section enable one to estimate also the coupling to a single SMM at any location with respect to the device. For a molecule placed in between the ground and central lines, we find that $g$ ranges between $100$ Hz and a few kHz, depending on the particular sample. Notice, however, that the magnetic field is enhanced, up to a factor $5$ or so, in narrow regions close to the edges of these lines (remember Fig. \ref{Bprofile}). Two distinctive aspects of SMMs, which are not easily found in other qubit realizations, is that they are sufficiently small, with lateral dimensions of the order of $1$ nm, to fit inside these regions and that they can be delivered from a solution with very high spatial accuracy by, e.g. using the tip of an atomic force microscope \cite{Martinez-Perez2011}. The magnetic field generated near the central line edges, thus also the coupling to molecules or molecular ensembles located near them, can be further enhanced by fabricating narrow constrictions. Superconducting circuits with dimensions well below $100$ nm can be fabricated, and even repaired, by either etching with a focussed ion beam or by using the same ion beam to induce the growth of a superconducting material from a gas precursor \cite{Martinez-Perez2009}. Provided that these constrictions are much shorter than the photon wave length, they are expected to have very little effect on the general resonator characteristics.

\begin{figure}
\centering
\includegraphics[width=0.8\textwidth]{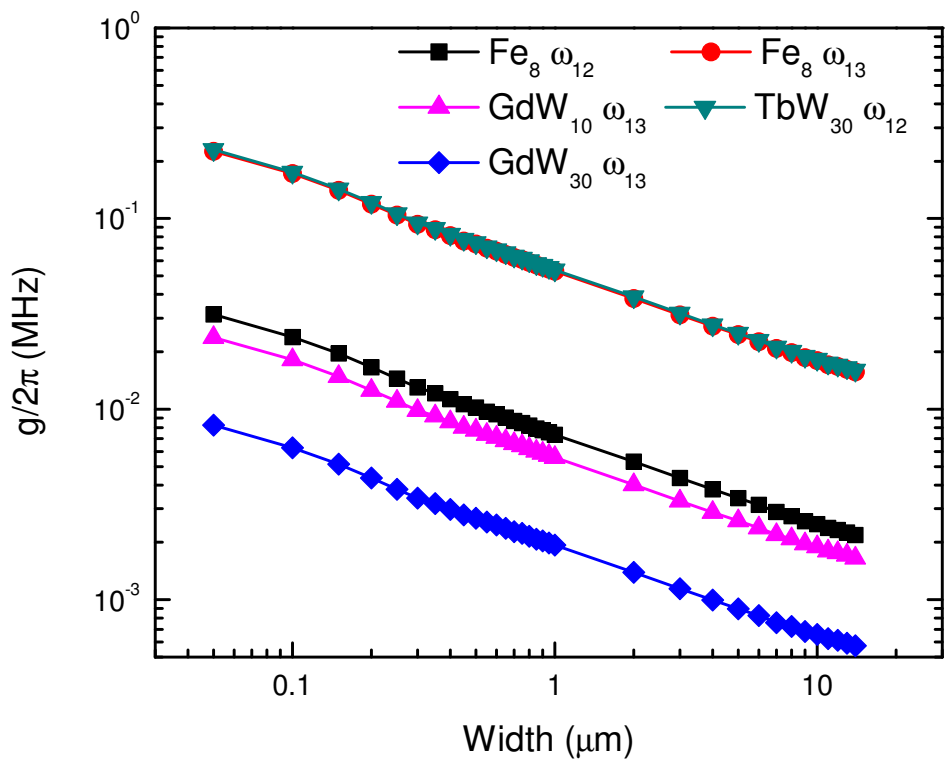}
\caption{Coupling of a single SMM to a CPW resonator as a function of centre line width. The SMM is located at the point of maximum field on the surface of the centre line.  For each sample, $\omega_{ij}$ denotes the transition used and the operating frequencies are detailed in Fig. \ref{gvsh}.}
\label{gvsw}
\end{figure}

In order to explore this possibility, we have repeated the above simulations for varying centre line widths, down to $50$ nm, while keeping current constant. We then evaluate the coupling to a single SMM located at the point of maximum field on the surface of the centre line and oriented in such a way as to maximize the transition matrix element. The results are shown in Fig. \ref{gvsw}. We see that reducing the width from $14\,\mu$m to $50$ nm can lead to enhancements of an order of magnitude in the coupling strength. Again, the dependence on the geometry is the same for all samples. The conclusion is that achieving strong coherent coupling of a single SMM (e.g. TbW$_{30}$) to such nanoresonators requires that that the decoherence time $T_{2}$ of an individual molecule grafted to a superconducting device can be made significantly longer than $10\,\mu$s. Despite the lack of $T_{2}$ data for truly isolated molecules, it seems that such coherence times can be reached under adequate conditions, i.e. for molecules having a very low concentration of nuclear spins \cite{Wedge2012}.

Before moving to the following section, it is worth mentioning here that the potential applications of superconducting resonators or transmission wave guides that maximize the magnetic coupling to very small spin ensembles, or eventually enable detecting single spins, extends well beyond the quantum information research field. For instance, these designs might contribute to the optimization of on-chip electron paramagnetic resonance spectrometers for the characterization of magnetic materials \cite{Clauss2013}.

\section{Coupling of SMMs to flux qubits}
\label{fluxqubits}

\subsection{Device description and parameters}

Flux qubits (FQs) are superconducting loops interrupted by, almost always, three  junctions \cite{Orlando1999}.
When half of a flux quanta passes through the loop, the two lowest eigenstates of the qubit Hamiltonian are symmetric and antisymmetric superpositions of counter propagating persistent currents.  Those states define the qubit states, and will be denoted as $\left\{ |\circlearrowleft\rangle,|\circlearrowright\rangle\right\}$. By changing the flux, the qubit is biased to one of those currents. Furthermore, any influence of higher excited levels can be safely neglected for standard qubit parameters. Therefore for our purposes the FQ can be modelled as a two level system:
\begin{equation}
\label{FQ}
H_{FQ}
= \frac{\epsilon}{2} \sigma_z
+
\frac{\Delta}{2} \sigma_x
\end{equation}
with $\Delta$ the qubit gap which lies in the GHz regime and $\epsilon = 2 I_p (\phi_{\rm ext} - \Phi_0/2)$ the  bias term associated with the external flux $\phi_{\rm ext}$ and $I_p$ the persistent current in the loop.  Here we have chosen the {\it physical} basis where the eigenstates are the clockwise and anticlockwise supercurrents: $\left\{ |\circlearrowleft\rangle,|\circlearrowright\rangle\right\} $.

FQs provide a  platform for hybrid structures because their ability to couple to magnetic moments through the field induced by the supercurrents in the loop. Previous studies have focused on considering the coupling to NV-centres \cite{Marcos2010}, with recent experimental realizations showing promising results \cite{Zhu2010}. Some applications of these hybrid structures to quantum information processing have been recently pointed out \cite{Hummer2012, Xiang2013, Lu2013}.

\subsection{Coherent Coupling to SMMs}
\label{fluxqubitsgeometry}

\begin{figure}
\centering
\includegraphics[width=0.7\textwidth]{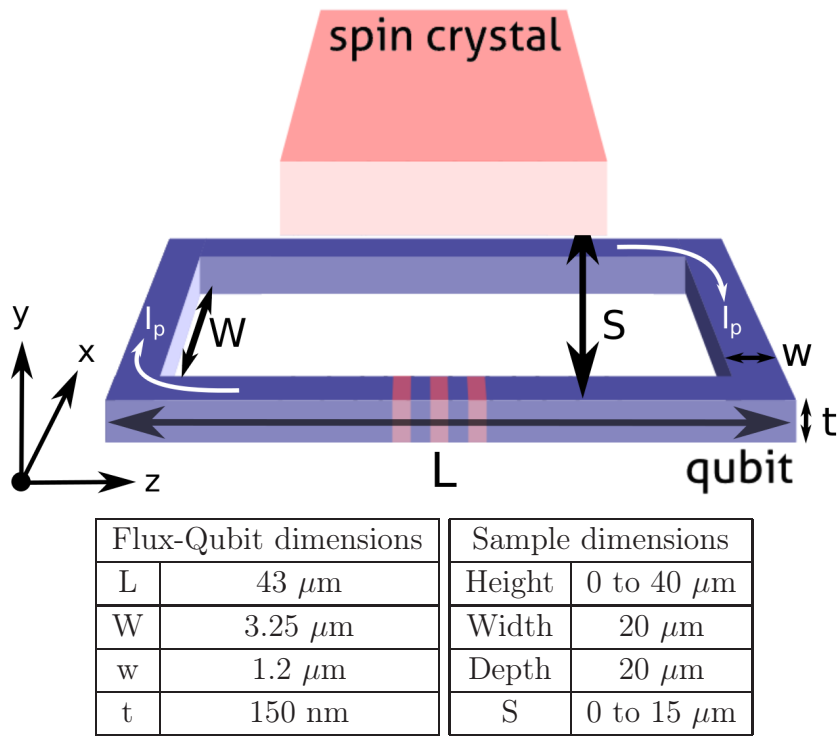}\\[2mm]
\begin{tabular}{|c|c|}
\hline
\multicolumn{2}{|c|}{Flux-Qubit dimensions} \\
\hline
L & 43 $\mu$m \\
\hline
W & 3.25 $\mu$m \\
\hline
w & 1.2 $\mu$m \\
\hline
t & 150 nm \\
\hline
\end{tabular}
\begin{tabular}{|c|c|}
\hline
\multicolumn{2}{|c|}{Sample dimensions} \\
\hline
Height & 0 to 40 $\mu$m \\
\hline
Width & 20 $\mu$m \\
\hline
Depth & 20 $\mu$m \\
\hline
S & 0 to 15 $\mu$m \\
\hline
\end{tabular}
\caption{Basic geometry and dimensions of the flux-qubit and the magnetic samples used for the calculations described in this work.  The flux-qubit dimensions resemble those from \cite{Zhu2011}}
\label{FQ-SMM}
\end{figure}

Within the state of the art of both qubit geometry and parameters the coupling to single spins is too weak to overcome the losses.
Therefore, in this section we will discuss the coupling between a flux qubit and a spin ensemble. A schematic representation of a possible layout is depicted in Fig. \ref{FQ-SMM}.

Following the same reasoning from the previous section, we can again arrive at a Rabi like model:
\begin{equation}
\label{rabi}
H = \frac{\Delta}{2} \sigma_x
+
\omega b^\dagger b
+
g \sigma_z (b^\dagger + b)
\end{equation}
where, for simplicity, we have chosen to be at the degeneracy point $\epsilon = 0$ [Cf. Eq. (\ref{FQ})].
The spin collective modes, $b, \; b^\dagger$,  were defined in equation (\ref{collective}) and the coupling in (\ref{coupling}).
In this case, the magnetic field, $b(\vec r_j)$ is generated by the circulating currents in the qubit [Cf. Eq. (\ref{coupling})].
The current operator can be written as,
\begin{equation}
I=\sum_{m,n=\circlearrowleft,\circlearrowright}|n\rangle\langle n|I|m\rangle\langle m|=I_{p}|\circlearrowleft\rangle\langle\circlearrowleft|-I_{p}|\circlearrowright\rangle\langle\circlearrowright|=I_{p}\sigma_{z}\, ,
\end{equation}
that justifies the coupling through $\sigma_z$. The magnetic field strength, $b(\vec r)$ entering in the formula for $g$, Eq. (\ref{coupling}), corresponds to the field generated in a loop with a circulating current $I_{\rm p}$.

To estimate this coupling, we again perform numerical simulations in Comsol Multiphysics assuming a superconducting loop with current $I_{\rm p}$.  As in the previous section, we simulate the field at a cross section at the centre of the flux qubit and choose a crystal size of about half the length of the flux qubit (see Fig. \ref{FQ-SMM}) to avoid edge effects. We also use the skin effect, as before, to simulate the superconducting current distribution (see equation (\ref{skineffect})). An example of the field distributions found is shown in figure \ref{Bprofile}.
We complement our numerical studies with an analytical approach.  In order to get a tractable and closed formula for the magnetic field generated, we approximate the qubit by two parallel counter currents, $I_p$.  This yields for the magnetic field:
\begin{eqnarray}
\mathbf{b} =\frac{\mu_{0}\, I_{p}}{2\,\pi}&\left(\frac{1}{\left(x+\nicefrac{w}{2}\right)^{2}+y^{2}}\left(\begin{array}{c}
-y\\
x+\nicefrac{w}{2} \\
0
\end{array}\right)\right.\nonumber \\
 & \left.-\frac{1}{\left(x-\nicefrac{w}{2}\right)^{2}+y^{2}}\left(\begin{array}{c}
-y\\
x-\nicefrac{w}{2}\\
0
\end{array}\right)\right)\,.
\label{analytical}
\end{eqnarray}

Both numerical and analytical estimates for $g$ are shown in Fig. \ref{gvsh_FQ}. We plot our results as a function of crystal height and as a function of the vertical separation between the crystal and the flux qubit (S in Fig. \ref{FQ-SMM}). As with the resonator, we observe a saturation of $g$ beyond a certain height.  This can also be understood by looking at the dependence on separation S. The field saturation occurs between $1$ and $10 \; \mu$m, i.e. when the magnetic field becomes negligible. We also observe that our simple analytical estimation closely reproduces the numerical results.

\begin{figure}
\centering
\includegraphics[width=0.65\textwidth]{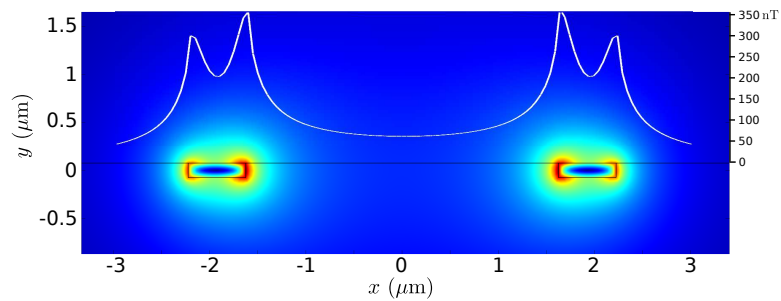}
\caption{Simulated field distribution on a flux qubit cross-section at the centre of the device.  The white profile is the field value calculated at a constant distance right at the surface of the superconducting regions.}
\label{Bprofile_FQ}
\end{figure}

\begin{figure}
\centering
\includegraphics[width=\textwidth]{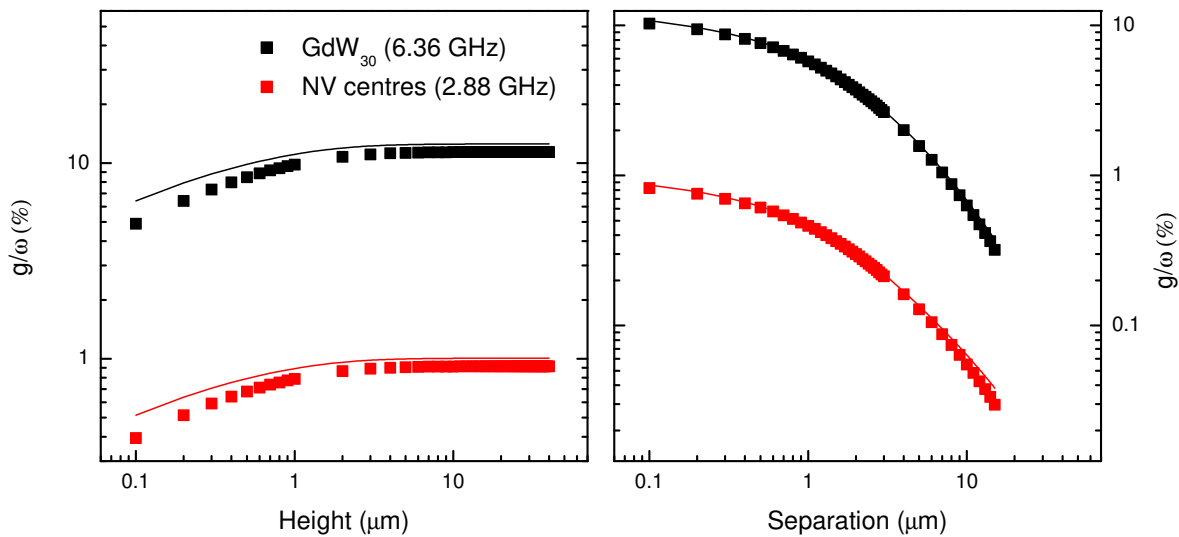}
\caption{Coupling of SMM crystals to a flux qubit as a function of crystal height (left) and as a function of the vertical (i.e. along $y$) separation between the crystal and the device, normalized by the qubit frequency $\omega$. Solid lines represent the analytical estimations that follow from Eq. (\ref{analytical}) while the dots are from the numerical simulation.}
\label{gvsh_FQ}
\end{figure}

In the case of FQs we have compared two species: GdW$_{30}$ and NV-centres, with zero field level splittings of $6.4 $ and $2.8 $ GHz, respectively. These transition frequencies lie in the range of  available qubit tunneling gaps. The achievement of strong coupling between NV-centres and a flux qubit has been recently reported in \cite{Zhu2010}. From the present results we conclude that, as we had anticipated, spin ensembles tend to couple more strongly to flux qubits than to resonators, Cf. Figs. \ref{gvsh} and \ref{gvsh_FQ}.  Also, as in the case of the resonators, the coupling to SMMs is stronger than that to NV-centres. This point is interesting since the qubit-ensemble coupling can reach up to $10 \% $ of the qubit natural frequency. For such a coupling strength, the qubit-spin ensemble system enters the so-called ultrastrong coupling limit \cite{Niemczyk2010,Forn-Diaz2010}. This means that the full model (\ref{rabi}) is needed to understand the physics. In the usual case of weaker coupling, one can rotate the qubit basis $\sigma_x \to \sigma_z$ and write the interaction within the Rotating Wave Approximation $g (\sigma^+ a + \sigma^- a^\dagger)$, with $\sigma_\pm = \sigma_x \pm i \sigma_y$. The latter approximation allows a perturbative treatment. Therefore, SMMs are candidates to observe analogues of light-matter interaction beyond perturbative treatments.  We finish by noting that the flux qubit parameters used here were taken from the experimental paper in Ref. \cite{Zhu2010}.  Further optimization of the parameters and of the flux qubit shape could yield even stronger couplings.

\section{Conclusions: why SMMs?}
\label{conclusions}

The results described in previous sections confirm that, because of their high spins and spin densities, SMMs have the potential to attain very high couplings with superconducting circuits. In addition, the great variety of magnetic molecules enables a vast choice of resonance frequencies. However, for many of the best-known SMMs, such as Fe$_{8}$ or Mn$_{12}$, the strong magnetic anisotropy introduces also some technical difficulties, i.e. the need to work at very high frequencies, above $100$ GHz, or the application of strong (above $2$ T) and very accurately alligned (within $0.5$ deg.) magnetic fields. For this reason, it will probably be more adequate to work with single ion magnets, i.e. molecules with just one magnetic ion. Compared with polynuclear clusters, these molecules have the advantage of being simpler, thus its physical response is easier to describe, offer a greater versatility for the modification of the spin Hamiltonian via the rational design of the local coordination shell surrounding the central magnetic ion, and can be made more robust against decoherence.

Yet, it seems natural to inquire whether SMMs might bring some new possibilities, not easily achievable with other spin systems. A first, {\em quantitative} answer to this question is given by the couplings of SMMs crystals to superconducting resonators and flux qubits that we find. In both cases, the collective coupling attains significant fractions, $\sim 10$ \%, of the natural circuit frequency, much larger than those observed so far for, e.g., NV centres in diamond. Under these conditions, the combined system enters the "ultra-strong" coupling limit, meaning that perturbative treatments are no longer applicable to describe the underlying physics. Of fundamental interest is the coupling of an SMM crystal to a flux qubit, because it is analogue to the light matter interaction in cavity QED \cite{Zhu2010}, the spin ensemble playing the role of an oscillator bath. Taking into account the vast ranges of parameters that can be explored (by e.g. varying the spin concentration or the energy gaps) this hybrid device can therefore be used to simulate the physics of open systems, help to understand and control the associated decoherence or develop noise resilient computation protocols.

From a more practical point of view, the attainment of strong coupling conditions might also confer to these systems interest as quantum memories \cite{Imamoglu2009,Wesenberg2009,Marcos2010}. A major difficulty arises though from the short lived spin coherence of these molecular systems. Decoherence times measured on SMMs crystals \cite{Takahashi2011} are still orders of magnitude shorter than those found for, e.g., NV centres \cite{Balasubramanian2009}. Therefore, SMMs cannot be considered for such applications unless coherence times are enhanced significantly. However, chemistry also provides suitable means to minimize the main sources of decoherence. For instance, isotopically purified molecules can be synthesized, in order to decrease the number of environmental nuclear spins \cite{Ardavan2007}. Also, decoherence caused by nuclear spin diffusion can be reduced by using sufficiently rigid ligand molecules \cite{Wedge2012}. Pairwise decoherence caused by dipolar interactions \cite{Morello2006} can be reduced by either dissolving the molecules in appropriate solvents \cite{Wedge2012,Ardavan2007,Schlegel2008,Bertaina2008} or by growing crystals in which a fraction of molecules is replaced by nonmagnetic ones \cite{Martinez-Perez2012}. Working with magnetically diluted samples has, however, a cost in terms of coupling. Therefore, a gain in performance (i.e. a net enhancement of $g_{N}T_{2}/2\pi$) can only be achieved provided that $T_{2}$ grows faster than $1/\sqrt(N)$, a condition that seems to hold in the very low temperature limit $k_{\rm B}T \ll \hbar \omega$, when magnon-mediated decoherence is expected to dominate \cite{Morello2006}. For a given spin density, the strength of dipolar interactions also decreases with $S$, thus it can be reduced by working with low-spin molecules, e.g. single ion magnets containing lighter lanthanide ions (Ce$^{3+}$,Sm$^{3+}$, or Gd$^{3+}$) or $S=1/2$ paramagnetic radicals \cite{Chiorescu2010,Abe2011} and Cr$_{7}$Ni molecular rings \cite{Troiani2005,Wedge2012,Ardavan2007}. The material of choice will therefore largely depend upon the attainment of an optimum tradeoff between maximizing $g_{N}$ and $T_{2}$.

But probably the main interest of SMMs is that they are also {\em qualitatively} different to most other spin systems in that they can be chemically engineered to fulfil very diverse functionalities. Restricting ourselves to the field of quantum information, magnetic molecules can be much more than single spin qubits \cite{Meier2003,Troiani2005}. Some molecular structures \cite{Timco2009,Candini2010,Aromi2012} embody several weakly coupled, or entangled, qubits which can provide realizations of elementary quantum gates  \cite{Luis2011} or act as quantum simulators \cite{Santini2011}. In addition, their multilevel magnetic energy structure can be used to encode multiple qubit states or even to perform quantum algorithms \cite{Leuenberger2001}. Coupling to quantum circuits can provide a method to experimentally realize these ambitious expectations, provided that one is able to strongly couple, thus coherently manipulate and read-out, individual molecules. In this respect, the fact that most SMMs are stable in solution opens the possibility to deposit them, in the form of monolayers or even individually, onto solid substrates \cite{Mannini2010} or at specific locations of a given device that maximize $g$ \cite{Martinez-Perez2011,Urdampilleta2011}. Our simulations show also that it is then possible to reach significantly larger couplings $g$, which can be further enhanced (up to $g/2\pi \sim 100-200$ kHz, see Fig. \ref{gvsw}) by the fabrication of narrow constrictions in the centre line of superconducting nanoresonators. These results suggest that the strong coupling limit is attainable for individual molecules, using state-of-the art technologies, provided that decoherence times can be made longer than $10-20\,\mu$s. Considering the available experimental evidences \cite{Wedge2012,Ardavan2007}, this limit, thus the realization of quantum technologies based on SMMs coupled to quantum circuits, seems definitely within reach.


\ack
The present work has been partly funded trough the Spanish MINECO (grants MAT2012-38318-C03, FIS2009-10061, FIS2011-25167 and FIS2012-33022), the European project PROMISCE,  CAM research consortium QUITEMAD S2009-ESP-1594 and the Gobierno de Arag\'on (projects MOLCHIP and FENOL).  Mark Jenkins acknowledges a JAE fellowship from CSIC.


\section*{References}


\end{document}